\documentclass[%
 reprint,
superscriptaddress,
 amsmath,amssymb,
 aps,
 prc,
]{revtex4-1}    

\usepackage[dvipdfmx]{graphicx}
\usepackage{dcolumn}
\usepackage{bm}
\usepackage{mathtools}
\usepackage{textcomp}
\usepackage{amsmath}
\usepackage{color}
\usepackage{epsfig}
\usepackage{epstopdf}
\usepackage{url}
\usepackage{hyperref}
\usepackage{CJK}
\usepackage[whole]{bxcjkjatype} 

\usepackage{setspace}
\usepackage[font=small,justification=raggedright]{caption}

\usepackage{color}

\begin{document}
\begin{CJK*}{UTF8}{}

\preprint{APS/123-QED}

\title{Electric monopole transition from the superdeformed band in $^{40}$Ca}

\author{E.~Ideguchi (井手口 栄治)}
\affiliation{Research Center for Nuclear Physics (RCNP), Osaka University, 10-1 Mihogaoka, Ibaraki, Osaka 567-0047, Japan}

\author{T.~Kib\'{e}di}
\affiliation{Department of Nuclear Physics and Accelerator Applications,
 Research School of Physics, The Australian National University, Canberra, ACT 2601, Australia}

\author{J.~T.~H.~Dowie}
\affiliation{Department of Nuclear Physics and Accelerator Applications,
 Research School of Physics, The Australian National University, Canberra, ACT 2601, Australia}

\author{T.~H.~Hoang}
\affiliation{Research Center for Nuclear Physics (RCNP), Osaka University, 10-1 Mihogaoka, Ibaraki, Osaka 567-0047, Japan}

\author{M.~Kumar~Raju}
\affiliation{Research Center for Nuclear Physics (RCNP), Osaka University, 10-1 Mihogaoka, Ibaraki, Osaka 567-0047, Japan}
\affiliation{Department of Physics, GITAM Institute of Science, GITAM University, Visakhapatnam-530045, India}

\author{N.~Aoi (青井 考)}
\affiliation{Research Center for Nuclear Physics (RCNP), Osaka University, 10-1 Mihogaoka, Ibaraki, Osaka 567-0047, Japan}

\author{A.~J.~Mitchell}
\affiliation{Department of Nuclear Physics and Accelerator Applications,
 Research School of Physics, The Australian National University, Canberra, ACT 2601, Australia}

\author{A.~E.~Stuchbery}
\affiliation{Department of Nuclear Physics and Accelerator Applications,
 Research School of Physics, The Australian National University, Canberra, ACT 2601, Australia}

\author{N.~Shimizu (清水 則孝)}
\affiliation{Center for Nuclear Study, The University of Tokyo, Hongo, Bunkyo-ku, Tokyo 113-0033, Japan}

\author{Y.~Utsuno (宇都野 穣)}
\affiliation{Advanced Science Research Center, Japan Atomic Energy Agency, Tokai, Ibaraki 319-1195, Japan}
\affiliation{Center for Nuclear Study, The University of Tokyo, Hongo, Bunkyo-ku, Tokyo 113-0033, Japan}

\author{A.~Akber}
\affiliation{Department of Nuclear Physics and Accelerator Applications,
 Research School of Physics, The Australian National University, Canberra, ACT 2601, Australia}

\author{L.~J.~Bignell}
\affiliation{Department of Nuclear Physics and Accelerator Applications,
 Research School of Physics, The Australian National University, Canberra, ACT 2601, Australia}

\author{B.~J.~Coombes}
\affiliation{Department of Nuclear Physics and Accelerator Applications,
 Research School of Physics, The Australian National University, Canberra, ACT 2601, Australia}

\author{T.~K.~Eriksen}
\affiliation{Department of Nuclear Physics and Accelerator Applications,
 Research School of Physics, The Australian National University, Canberra, ACT 2601, Australia}

\author{T.~J.~Gray}
\affiliation{Department of Nuclear Physics and Accelerator Applications,
 Research School of Physics, The Australian National University, Canberra, ACT 2601, Australia}

\author{G.~J.~Lane}
\affiliation{Department of Nuclear Physics and Accelerator Applications,
 Research School of Physics, The Australian National University, Canberra, ACT 2601, Australia}

\author{B.~P.~McCormick}
\affiliation{Department of Nuclear Physics and Accelerator Applications,
 Research School of Physics, The Australian National University, Canberra, ACT 2601, Australia}

\date{\today} 

\begin{abstract}
The electric monopole ($E0$) transition strength $\rho^2$ for the
 transition connecting the third 0$^+$ level, a ``superdeformed'' band
 head, to the ``spherical'' 0$^+$ ground state in doubly magic $^{40}$Ca
 has been determined 
via $e^+e^-$ pair-conversion spectroscopy. 
The measured value, $\rho^2(E0; 0^+_3 \to 0^+_1)~=~2.3(5)\times10^{-3}$, 
is the smallest $\rho^2(E0; 0^+ \to 0^+)$ found in $A<50$ nuclei. 
In contrast, the $E0$ transition strength to the ground state 
 observed from the second 0$^+$ state, a band head of ``normal''
 deformation, 
is an order of magnitude larger, $\rho^2(E0; 0^+_2 \to
 0^+_1)~=~25.9(16)\times~10^{-3}$, 
which shows significant mixing between these two states. 
Large-Scale Shell Model (LSSM) calculations were performed to understand 
the microscopic structure of the excited states, and the configuration 
mixing between them; experimental $\rho^2$ values in $^{40}$Ca and 
neighboring isotopes were well reproduced by the LSSM calculations. 
The unusually
 small $\rho^2(E0; 0^+_3 \to 0^+_1)$ value is 
due to destructive interference in the mixing of shape-coexisting
 structures, which are based on several different
 multiparticle-multihole 
excitations. 
This observation goes beyond the usual treatment of $E0$ strengths, 
where two-state shape mixing cannot result in destructive interference.
\end{abstract}

\pacs{Valid PACS appear here}

\maketitle
\end{CJK*}


Shape coexistence is a unique feature of self-bound, finite, quantum
many-body systems in which two or more different shapes emerge at
similar excitation energies. This phenomenon is known to manifest in
small atomic clusters \cite{Hor06} and molecules \cite{Lag04}. 
It also appears to be ubiquitous in atomic nuclei \cite{hey11}. 
The shape of a nucleus is
determined by the mean field generated by its constituent protons and
neutrons. It is influenced by the number of nucleons, energy level
density at the Fermi surface, an attractive residual interaction, and
the principle that nuclear shapes may change to lower the energy of the
system. A crucial difference between the atomic nucleus and more
macroscopic systems is that quantum-mechanical tunnelling causes
coexisting shapes to mix; each observed state is a superposition of
configurations that correspond to the various shapes. In particular,
shape coexistence
occurs near closed-shell nuclei, where 
it 
is based on competition between the stabilizing effect of nucleon shell
closures to retain a spherical shape, and the proton-neutron residual
interaction which drives deformation via multiparticle-multihole
excitations \cite{hey11}.

With the aid of modern radioactive-ion-beam facilities, shape
coexistence has also been identified in nuclei far from the valley of
$\beta$ stability. This potentially affects locations of nuclear drip
lines and waiting points that influence competition between $\beta$
decay and neutron capture in cosmic nucleosynthesis. 
Experimental indications of shape coexistence have been reported in the
double-closed-shell, neutron-rich nucleus $^{78}$Ni \cite{Tan19}.
It can also be closely associated with the break
down of familiar shell structures, such as suppression of the $N=20$
shell gap in neutron-rich $^{32}$Mg \cite{Gui84,Mot95}. The 0$^+$ ground
state, with very large quadrupole deformation parameter
$\beta_2\approx0.6$, is supposed to coexist with a near-spherical
first-excited 0$^+$ level \cite{Wim10}. Shape coexistence is generally
discussed in terms of a two-state mixing model \cite{hey11}. However, a
two-state analysis of $^{32}$Mg questioned the aforementioned
interpretation of a deformed ground state and spherical excited state
\cite{For11,For12}. 
If three-level mixing is applied in $^{32}$Mg \cite{Mac16,Mac17}, 
$B(E2)$ values, level energies, and 
transfer cross sections can be successfully explained, 
but the 0$_3^+$ level in this nucleus has not been experimentally
identified yet. 
More recently, the insufficiency of the two-state mixing model applied
to the 2$^+$ states in $^{42}$Ca was also discussed \cite{Had18}. 
Doubly magic $^{40}$Ca exhibits three distinct forms of quadrupole deformation: 
``spherical'', ``normal deformation'' (ND), and ``superdeformation'' (SD). 
Therefore, it provides a rare opportunity to study mixing effects between 
multiple configurations within a single nuclide.
The first-excited, 0$_2^+$ state at 3.35~MeV is the head of a ND rotational 
band. 
The second-excited, 0$_3^+$ level at 5.21~MeV is the head of a SD band \cite{ide01,chi03}. 
In addition, the 2$_2^+$ level at 5.24~MeV is interpreted as a member of 
a $K = 2$, four-particle four-hole (4p-4h) band \cite{ger69}.
Emergence of various structures in low-lying levels in $^{40}$Ca indicates 
shape coexistence. 
The main configurations for the ND and SD structures are 
4p-4h, and 8p-8h excitations across the $N,Z = 20$ shell gap, 
respectively \cite{For79,Mid72,Boh77,Ger67,ger69,cau07}. 
The transition quadrupole moments for the 
low-spin and high-spin
part 
of the SD band were reported to have a significant difference, 
indicating 
the mixing of lower-spin states with a less deformed configuration\cite{chi03}. 

Furthermore, there %
is another unique feature of the SD band in $^{40}$Ca. 
Although SD nuclei are reported in several mass regions \cite{sin02}, 
SD band heads with $J^{\pi} = 0^+$ are only observed in the $A=40$
\cite{sve00,ide10} and fission-isomer \cite{sin02} regions. 
This makes it difficult to study their properties, such as mixing with
less-deformed configurations, in detail. 
Therefore, $^{40}$Ca provides a unique testing ground in which the
electric monopole ($E0$) transition strength $\rho^2(E0; 0^+ \to 0^+)$
between an SD band head and spherical ground state can be studied as a
direct probe of shape mixing \cite{woo99}. 

 The $\rho^2(E0; 0^+_3 \to 0^+_1)$ in $^{40}$Ca was previously
 investigated via the $^{40}$Ca(p,p$^\prime$) reaction by measuring
 $e^+e^-$ pair decay with a
 plastic-scintillator pair spectrometer \cite{ulr77}.
 However, insufficient energy resolution meant that this state at
 5.21~MeV was not resolved from the 2$_2^+$ state at 5.25~MeV, and an
 upper limit of
 $\rho^2(E0; 0^+_3 \to 0^+_1)<0.06$ was deduced.
 In order to accurately determine the value of $\rho^2(E0; 0^+_3 \to
 0^+_1)$ and understand the properties of the SD band, measuring the
 $E0$ transition with higher energy
 resolution and low-background conditions is critical.


 This Letter reports on a new study of excited states in $^{40}$Ca following proton inelastic
 scattering from a self-supporting, 1.5-mg/cm$^2$ thick, natural Ca target.
 Proton beams were delivered by the 14UD Pelletron tandem accelerator of the Heavy Ion
 Accelerator Facility at the Australian National University.
 The optimum beam energy to populate the $0_3^+$ state, 8.6~MeV, was determined by scanning
 the beam energy and comparing relative yields of the 1.308-MeV, $0_3^+
 \to 2_1^+$ and 5.249-MeV, $2_2^+ \to 0_1^+$  $\gamma$-ray transitions.

 Electron-positron ($e^+e^-$) pair decays from excited states were measured by the
 superconducting solenoid, Super-e spectrometer \cite{kib90,Eri20,Dow20_PLB}.
 The solenoid axis is perpendicular to the beam axis.
 The  $e^+e^-$ pairs emitted from the target are transported by the magnetic field to
 the Miel detector, an array of six 9-mm thick Si(Li) crystals.
 Two axial, Heavymet baffles are mounted between the target and detector to block
 $\gamma$ rays and X~rays, scattered beam particles, and secondary electrons.
 In this arrangement, $e^+e^-$ pairs of nearly equal energy can reach the detector.
 The thickness of the segments allows for full absorption of pair decays formed from transition
 energies up to 8~MeV.
 The efficiency of the pair-conversion spectrometer was derived from Monte Carlo simulations
 that consider the magnetic field, energy, and angular correlation of emitted $e^+e^-$ pairs.
 Spin alignment in the reaction and consequent angular distributions of
 $e^+e^-$-pair emission were taken into account \cite{Eri20}.
%

\begin{figure}[t]
\begin{center}
\includegraphics[width=\columnwidth]{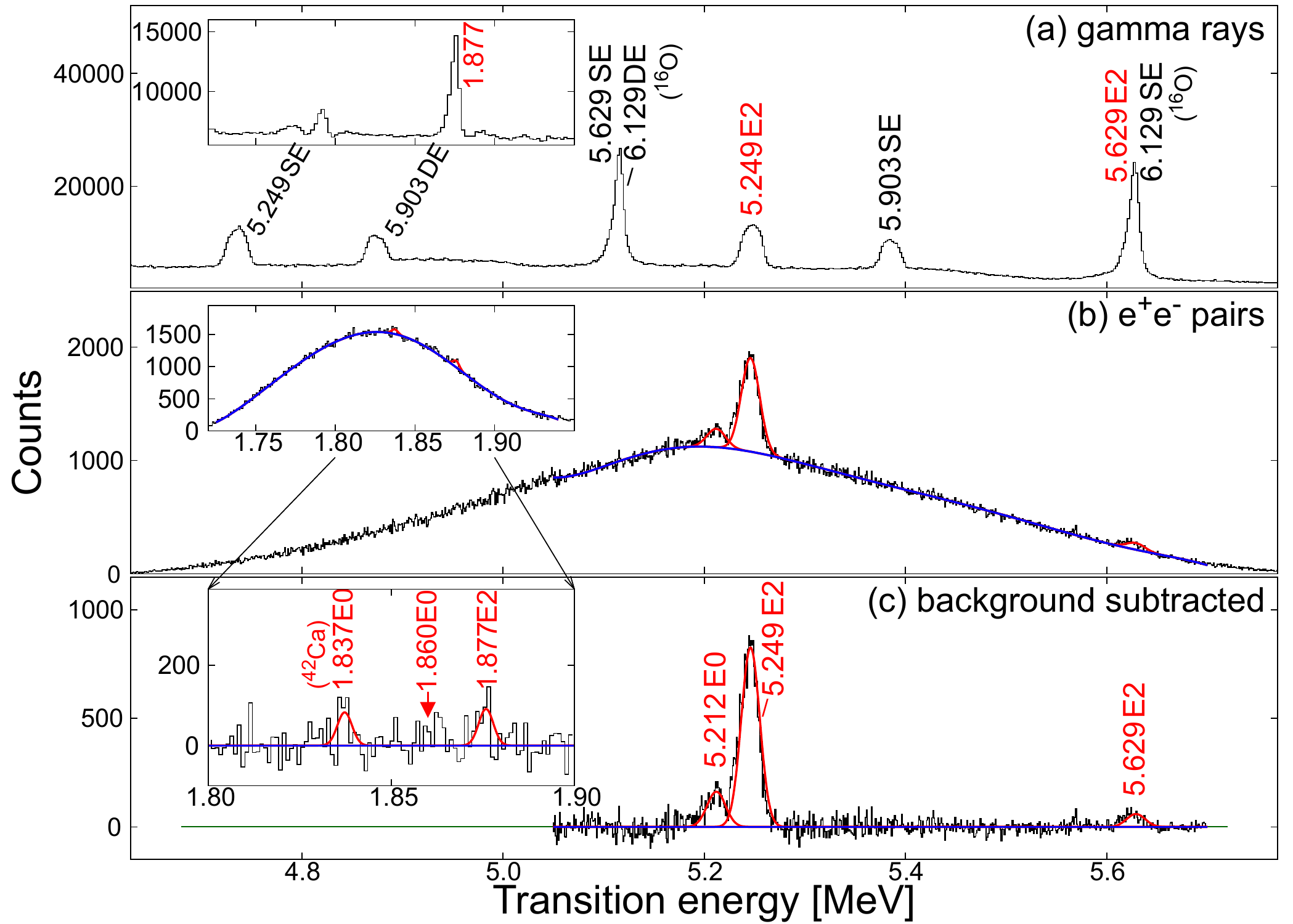}
\end{center}
\vspace*{-0.5cm}
\caption{\label{fig:spectra} (a) The $\gamma$-ray energy spectrum measured by the HPGe detector;
 (b) $e^+e^-$ pair coincidence spectrum;
 (c) and $e^+e^-$ spectrum after background subtraction.
 Peaks at 5.212, 5.249, and 5.629 MeV are labelled, as are single-escape (SE) and
 double-escape (DE) contaminant peaks.
 }
\end{figure}

\begin{figure}[t]
\begin{center}
\includegraphics[width=\columnwidth]{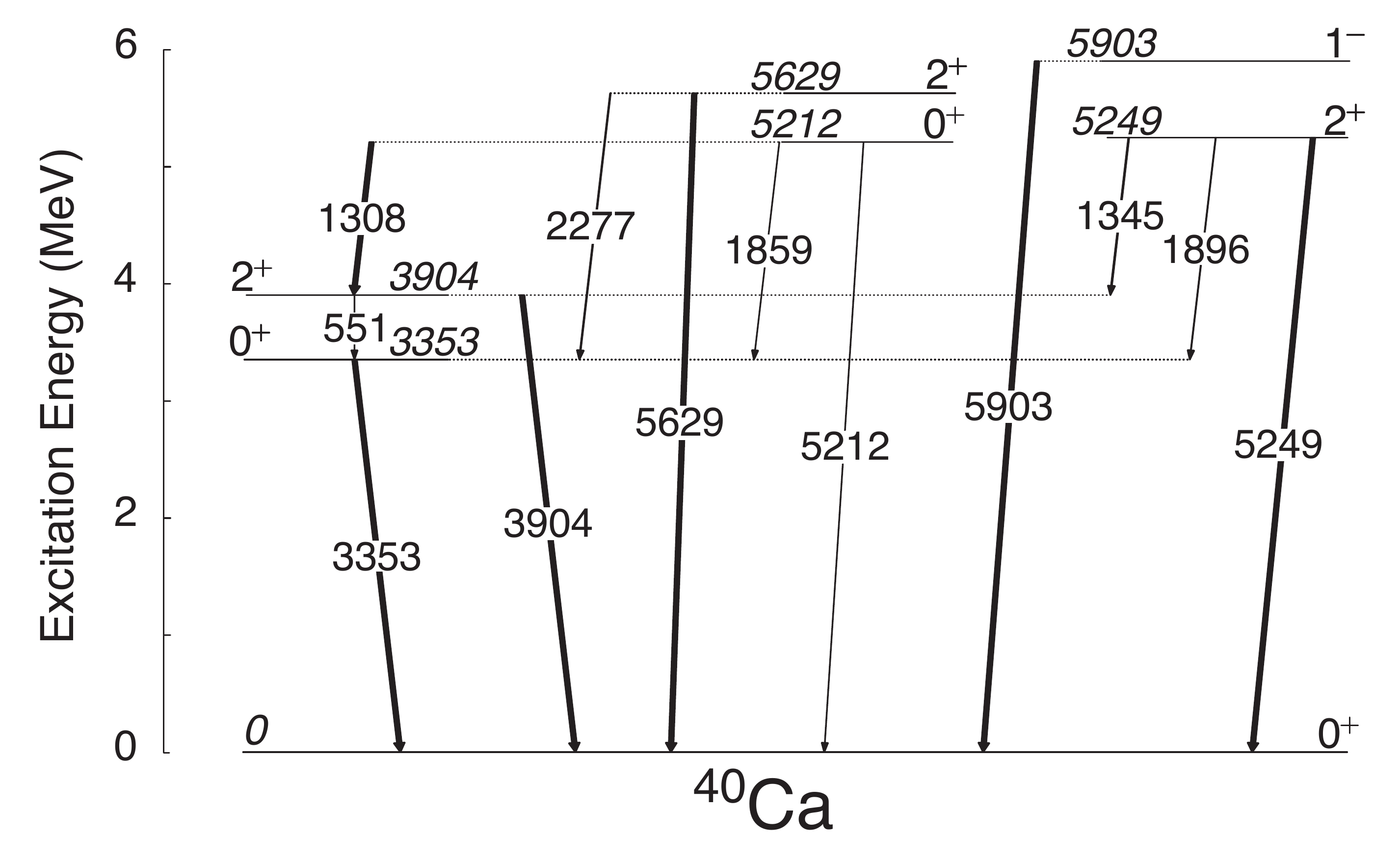}
\end{center}
\vspace*{-0.7cm}
\caption{\label{fig:levelscheme} Partial level scheme of $^{40}$Ca \cite{nds140}.}
\end{figure}

 A High-Purity Germanium (HPGe) detector was placed 1.5~m from the target and 135$^{\circ}$
 relative to the beam axis to simultaneously detect $\gamma$ rays emitted in the reactions.
 The relative $\gamma$-ray detection efficiency was measured using a $^{56}$Co source.
 Energies and detection times from the six Miel segments, energies from the HPGe detector,
 and the magnetic field values, were stored event-by-event and sorted offline.


 The $\gamma$-ray and $e^+e^-$ coincidence spectra analyzed to extract the $\rho ^2(E0)$ values
 are shown in Figure \ref{fig:spectra}.
 Gamma-ray photo peaks, as well as single- and double-escape peaks associated with the
 5.249- $(2^+_2 \to 0^+_1)$, 5.629- $(2^+_3 \to 0^+_1)$,
 and 5.903-MeV $(1^-_1 \to 0^+_1)$ transitions of $^{40}$Ca, are evident.
 The half-lives of these states are 83$^{+11}_{-9}$, 40(15), and 15.8(22)~fs,
 respectively \cite{nds140}.
 Since the stopping time of $^{40}$Ca recoils in the target is approximately 0.3~ps, these
 $\gamma$ rays are emitted while the nucleus is in motion and the peaks are consequently Doppler
 broadened.
The $\gamma$-ray energy spectrum corrected for Doppler shift
is presented in Fig.~\ref{fig:spectra}(a).
 Escape peaks of the 6.129-MeV $\gamma$~ray in $^{16}$O \cite{til93},
 from oxidation of the Ca target, are also visible.
 Since the half-life of this state is 18.4~ps, these $\gamma$~rays are emitted after stopping
 and the associated peaks are sharp.

 The $e^+e^-$ pair spectrum is shown in Fig.~\ref{fig:spectra}(b) with the energy axis shifted
 by 1.022~MeV to correspond to the associated transition energies.
 A background that was 
parametrized by a polynomial function was subtracted, giving the $e^+e^-$
pair spectrum presented in Fig.~\ref{fig:spectra}(c).
 There is a peak at 5.212~MeV with no associated $\gamma$~ray; this is the $E0$, $0^+_3 \to 0^+_1$
 transition.
 The only other peaks observed in panels (b) and (c) are the 5.249- and 5.629-MeV, $E2$
 transitions from the 2$_2^+$ and 2$_3^+$ levels to the ground state.

 Extraction of $\rho^2$ values was based on analysis of $\gamma$-ray and
 $e^+e^-$ spectral yields.
 The term $\rho^2$ is related to the measured $E0$ transition rate,
 $1/\tau(E0)$, by \cite{woo99}:
\begin{equation}
\frac{1}{\tau(E0)} = \rho^2(\Omega_K+\Omega_{L_1}+...+\Omega_{\pi}) \, ,
\end{equation}
\noindent
 where the $\Omega_j$ terms are electronic factors associated with atomic shells
 ($K, L_1, ...$). 
For both electron and $e^+e^-$ pair conversion ($\pi$) they depend on the atomic number
 and transition energy; numerical evaluations %
are available \cite{kib08,do20}.
 The electron conversion contributions to the $E0$ transitions considered here
 are $<$0.6\% ($0_2^+ \to 0_1^+$ and $0_3^+ \to 0_1^+$) and 11\%
 ($0_3^+~\to~0_2^+$). 
The relevant values of $\Omega_{\pi}$ are included in Table~\ref{table1}.

%
%

\begin{table}[t]
\begin{center}
\caption{\label{table1} Summary of $\rho^2(E0; 0^+ \to 0^+)$ values in $^{40}$Ca.}
\renewcommand{\arraystretch}{1.5}
\begin{tabular*}{0.48\textwidth}{@{\extracolsep{\fill} }ccccc}
\toprule
Transition & Energy & $\tau (E0)$ & $\Omega_{\pi}(E0)$ & $\rho^2\times10^3$ \\
		&  (MeV)  &  (ns) 	    &  	\cite{do20}				 &		 \\
\hline
0$_2^+ \rightarrow$ 0$_1^+$ & 
3.353                       & 
3.13(12)                    & 
1.24(6) $\times 10^{10}$    & 
25.9(16)                    \\
0$_3^+ \rightarrow$ 0$_1^+$ & 
5.212                       & 
3.2(7)                      & 
1.36(7) $\times 10^{11}$    & 
2.3(5)                      \\
0$_3^+ \rightarrow$ 0$_2^+$ & 
1.859                       & 
$>$76                       &
2.79(14) $\times 10^{8}$    &
$<$45                       \\
\hline
\hline
\end{tabular*}
\end{center}
\end{table}

 The mean lifetime of the $0_3^+$ state, $\tau = 1.47(30)$~ps, is known from Doppler-shift
 attenuation measurements \cite{pol69}.
 Figure~\ref{fig:levelscheme} shows the partial level scheme of $^{40}$Ca relevant to the present
 analysis \cite{nds140}; the $0_3^+$ state decays by a 1.308-MeV, $E2$ $\gamma$-ray transition,
 or by 1.859- or 5.212-MeV, $E0$ transitions.
 While the 1.859-MeV, $0_3^+ \to 0_2^+$ $E0$ component was sought by optimizing the Super-e
 magnetic field for that transition energy, 
 a clear peak was not observed above background (see Fig.~\ref{fig:spectra}(c)); however,
 an upper limit was established for the branching ratio.
 Therefore, using the measured $\gamma$-ray and $e^+e^-$ pair intensities and theoretical
 pair conversion coefficients from BrIcc \cite{kib08}, all decay branches from the
 $0_3^+$ state were determined.
 A value of $\rho^2(E0; 0_3^+ \to 0_1^+) =2.3 (5) \times 10^{-3}$ was
 extracted.
 An upper-limit of $\rho^2(E0; 0_3^+ \to 0_2^+) < 4.5\times10^{-2}$ was also obtained for
 the 1.859-MeV transition.
 The significantly smaller value of $\Omega_{\pi}(E0)$ explains why this transition
 was not observed directly in the experiment.

\begin{figure}[t]
\begin{center}
\includegraphics[width=\columnwidth]{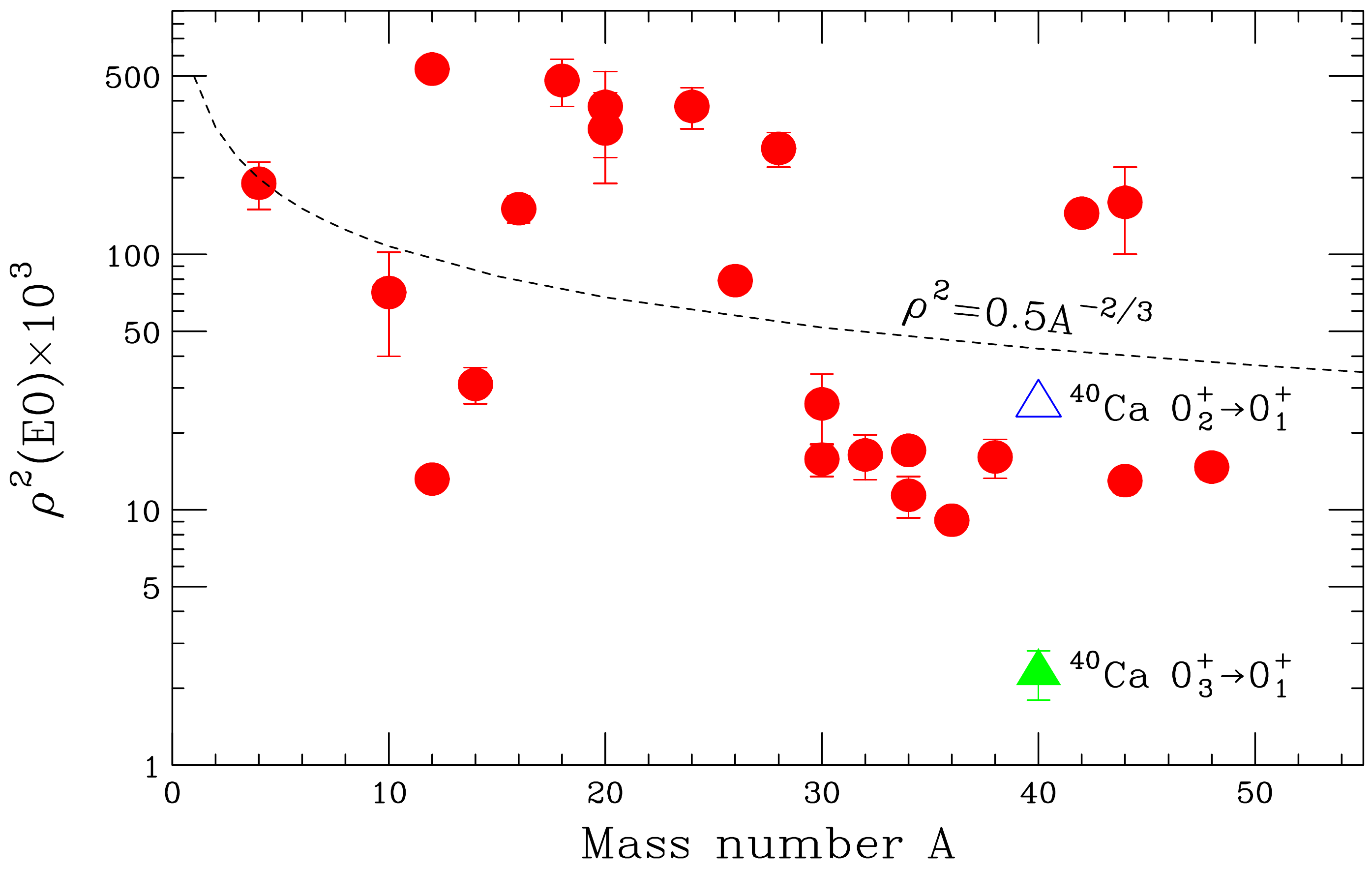}
\end{center}
\vspace*{-0.3cm}
\caption{\label{fig:rho_sys50} Systematics of $\rho^2(E0; 0^+\to 0^+_{1})$ values
 of even-even, $A<50$ nuclei (filled circles) \cite{kib22}.
 The $\rho^2(E0; 0^+_{3,2} \to 0^+_1)$ in $^{40}$Ca from this work are shown as
 open and filled triangles.}
\end{figure}

 Since the $0_2^+$ level is the lowest excited state and, consequently, has a single $E0$
 decay branch to the ground state, evaluation of
 $\rho^2(E0; 0_2^+~\to~0_1^+) = 25.9(16) \times~10^{-3}$ from the known lifetime was
 less complex.
 This $\rho^2$ value is large, indicating significant shape differences
 and mixing between the two states.
 The $E0$ transition strengths from our 
work are summarized in Table \ref{table1}.
 Systematic behavior of $\rho^2(E0; 0^+ \to 0^+_{\rm{1}})$ values for even-even nuclei
 with $A < 50$ \cite{kib22} is presented in Figure~\ref{fig:rho_sys50}.
 The dashed line corresponds to $\rho^2(E0)=0.5A^{-2/3}$ \cite{woo99}.
 The $\rho^2(E0; 0_2^+ \to 0_1^+)$ value measured in the present 
work shows good agreement with
 the systematic trend.
 Conversely, $\rho^2(E0; 0_3^+ \to 0_1^+)$ is significantly smaller than the trendline,
 and all of the other experimental 
values.

 In general, the $E0$ transition strength between states in a nucleus gives a direct measure of
 shape mixing between them \cite{Ald72}.
 The associated mixing amplitude $\alpha$ is usually estimated by considering a simple
 two-state model with spherical and deformed wave functions \cite{woo99}.
 In such a case, $\rho^2(E0)$ is related to $\alpha$ and the difference in
square of quadrupole deformation parameters, $\Delta(\beta_2^2)$, by: \vspace*{-0.2cm}
\begin{equation}
\label{eq:two_state}
\rho^2(E0)=\left ( \frac{3}{4\pi } Z\right )^2\alpha^2(1-\alpha^2)\left [ \Delta (\beta_2^2) \right ]^2 \, .
\end{equation}

Using the measured $\rho^2(E0)$ %
and the reported $\beta_2$ values of 0.27(5) and 0.59$_{-0.07}^{+0.11}$ for
the 0$_2^+$ and 0$_3^+$ states, respectively \cite{ide01},
$\alpha_{2 \to 1} = 0.55(37)$ and
$\alpha_{3 \to 1} = 2.9^{+1.1}_{-0.8} \times 10^{-2}$ 
were determined for the $0_2^+ \rightarrow 0_1^+$ and $0_3^+ \rightarrow
0_1^+$ transitions.
Here, the $\beta_2(0_1^+) = 0$ \cite{gar16} was assumed. 
Within the two-state model, the small mixing for the $0_3^+ \rightarrow 0_1^+$ transition
 implies a very small overlap of wave functions between the spherical ground state and SD
 band head.

 Large-Scale Shell-Model (LSSM) calculations were performed to gain a microscopic understanding
 of the measured $\rho^2(E0)$ values and the structure of $^{40}$Ca.
 Assuming an inert $^{16}$O core, the valence model space included the full $sd$ shell
 ($0d_{5/2}, 1s_{1/2}, 0d_{3/2}$) and restricted $fp$ shell ($0f_{7/2}, 1p_{3/2}$) orbitals.
 The calculations used $E0$ effective charges of $(e_p,e_n)=(1.8,0.8)e$, and harmonic-oscillator
 (HO) single-particle wave functions with %
$\hbar\omega=45A^{-1/3}-25A^{-2/3}$~MeV.

 The effective interaction was designed to describe multiparticle-multihole ($mp$-$mh$)
 excitations across the $N=Z=20$ shell gap by tuning the SDPF-M interaction \cite{uts99,uts04}
 within the model space, which was truncated to include $m\le 10$.
 The tuning was carried out so that one-neutron separation energies of $^{40,41}$Ca and
 one-$\alpha$ separation energies of $^{40}$Ca and $^{44}$Ti were reproduced with the
 many-body correlations that were included.
 Recently, this interaction was successfully applied to high-spin states in $^{35}$S \cite{go21}.

 Calculated $mp$-$mh$ probabilities for $0^+_{1,2,3}$ states of $^{40}$Ca are listed in
 Table~\ref{tab:ca40mpmh}.
 The dominance of $m=0, 4, 8$ configurations in $0^+_{1,2,3}$ is clearly observed; 
 this structure is consistent with Ref.~\cite{cau07}.
 There is also considerable mixing, especially with neighboring values of $m$.

\begin{table}[t]
\begin{center}
\caption{\label{tab:ca40mpmh} Calculated excitation energies and $mp$-$mh$ probabilities of the $0^+_{1,2,3}$ states in $^{40}$Ca.}
\renewcommand{\arraystretch}{1.5}
\begin{tabular*}{0.48\textwidth}{@{\extracolsep{\fill} }cccccccc}
\toprule
  & $E_x$ (MeV) & $0p$-$0h$ & $2p$-$2h$ & $4p$-$4h$ & $6p$-$6h$ & $8p$-$8h$ & $10p$-$10h$ \\ \hline
\hline
$0^+_{\textrm{1}}$ & 0     & 0.46 & 0.39 & 0.13 & 0.02 & 0.00 & 0.00 \\
$0^+_{\textrm{2}}$ & 2.81  & 0.04 & 0.03 & 0.63 & 0.26 & 0.03 & 0.00 \\
$0^+_{\textrm{3}}$ & 6.37  & 0.02 & 0.07 & 0.11 & 0.23 & 0.56 & 0.01 \\
\hline
\hline
\end{tabular*}
\end{center}
\end{table}

\begin{figure}[t]
\includegraphics[width=\columnwidth]{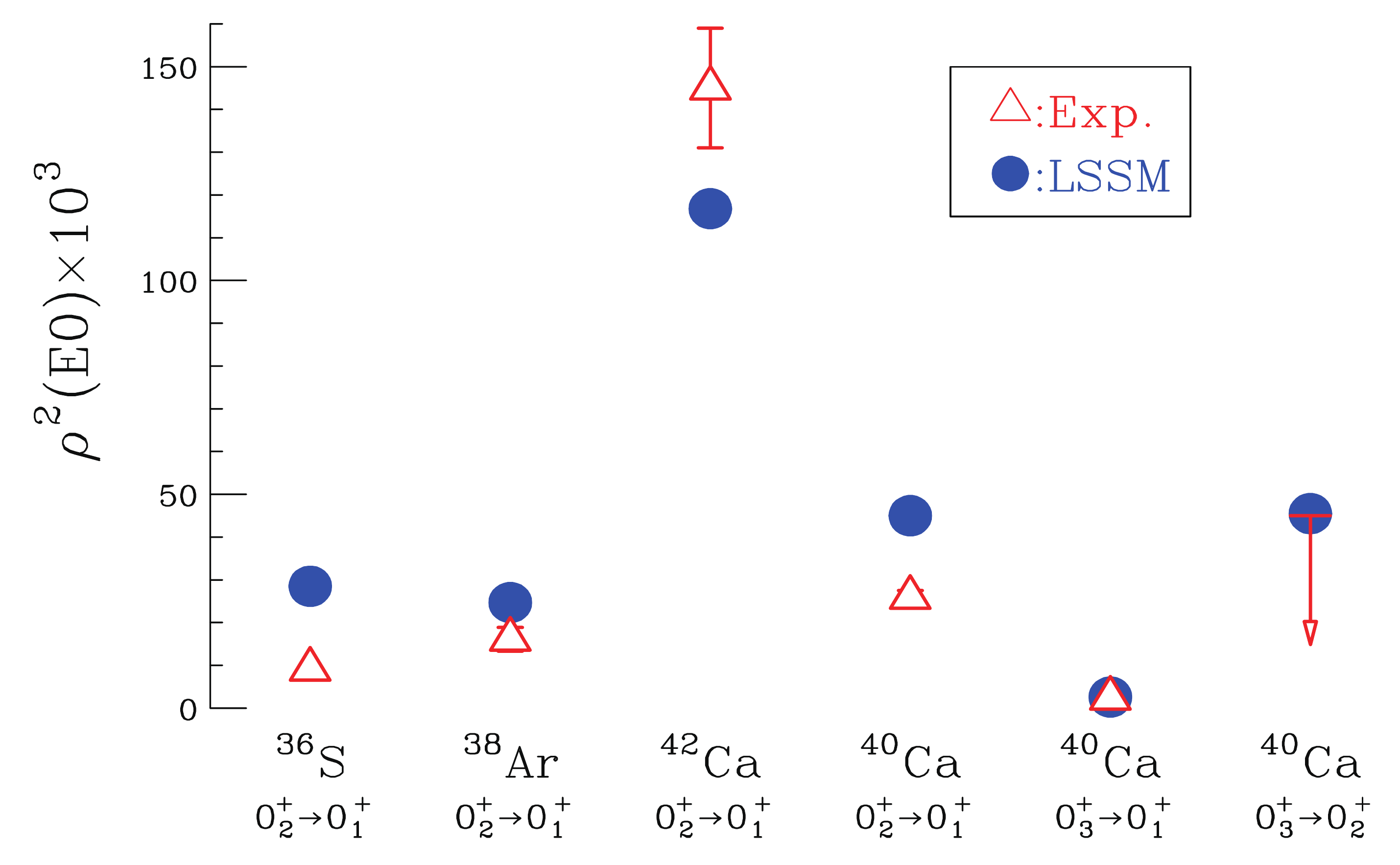}
\vspace*{-0.0cm}
\caption{\label{fig:rho2_lssm} Systematics of $\rho^2(E0)$ values for
 $^{36}$S, $^{38}$Ar, $^{42}$Ca and $^{40}$Ca. Open triangles are
 experimental values from Refs.~\cite{kib22,oln71} and this work; filled
 circles are LSSM calculations.}
\label{fig:e0_a40}
\end{figure}

Figure \ref{fig:rho2_lssm} shows $\rho^2(E0)$ values for $A\approx 40$ nuclei whose excited $0^+$ states are
dominated by $mp$-$mh$ excitations.
Data for $^{36}$S, $^{38}$Ar, $^{42}$Ca \cite{kib22,oln71} and $^{40}$Ca (this
work) are plotted with the corresponding LSSM calculations. Overall, the
theory reproduces the experimental data fairly well; in particular,
excellent agreement was found for the $0_3^+ \to 0_1^+$ in $^{40}$Ca
despite its value being lowest among the $Z<50$ nuclei. The upper limit
value for the $0_3^+ \to 0_2^+$ transition is also consistent with the
LSSM result.

 All of the ground states considered in Fig.~\ref{fig:e0_a40} are near spherical, whereas the $0^+_2$
 states in $^{36}$S, $^{38}$Ar and $^{40}$Ca are considered to have normal deformations \cite{wood92}.
 With similar $\rho^2(E0)\times 10^{3} \approx$ 10 to 20 values, these $0^+_2 \to 0^+_1$ transitions
 are in accordance with the two-state analysis of Eq.~(\ref{eq:two_state}) when similar mixing
 amplitudes are employed.
 However, the $0^+_2$ state in $^{42}$Ca \cite{had16} and the $0^+_3$ state in $^{40}$Ca \cite{ide01}
 are strongly deformed, yet their $\rho^2$ values for the transition to the ground state are
 completely different.
 This unexpected observation stimulates a detailed analysis of the role of mixing, which is described
 below.

In the present calculation, HO wave functions were used in a model space
that prohibited single-nucleon excitations of $2\hbar\omega$, such as
$0s \to 1s0d$. A many-body state, $|\Psi_i \rangle$, may be decomposed
into its $mp$-$mh$ components, $|\Psi_i\rangle = \sum_{m} |\Psi_i
(m)\rangle$, such that: \\[-0.7cm]

\begin{equation}
\langle \Psi_{f} | r^2 | \Psi_i \rangle = b^2\sum_m w(m) \xi_{if}(m) \, ,\\[-0.2cm]
\end{equation}

\noindent
where $b=\sqrt{\hbar/(M\omega)}$ is the HO length parameter, $w(m) = m +
c(^{40}\rm{Ca})$ are the weighting factors for each value of $m$ with
$c(^{40}\rm{Ca}) = 120$, and $\xi_{if}(m) =~\langle \Psi_{f}(m) |
\Psi_i (m) \rangle$ are the overlap factors between states under
inspection. Here, only cases where initial and final states, denoted $i$
and $f$, with $0^+$ assignments are considered. For $f\ne i$, $\sum_m
\xi_{if}(m)=0$ must also be satisfied from the imposed orthogonality
condition, $\langle \Psi_{f} | \Psi_i \rangle = \delta_{if}$.

Since $N = Z = 20$ in $^{40}$Ca, $T=0$ for all $0^+$ states and only
isoscalar transitions contribute. Hence, the $E0$ matrix element is
given by: \\[-0.8cm]

\begin{align}
M(E0;i \to f) &= \sum_m M_{if}(m), \nonumber \\
		      &= e_{\textrm{IS}} \times b^2 \cdot \sum_m (w(m) \cdot \xi_{if}(m)),
\end{align}

\noindent
where $e_{\textrm{IS}}=(e_p+e_n)/2$. The individual $M_{if}(m)$ terms
are, therefore, described by a weight factor $w(m)$ and an overlap
factor $\xi_{if}(m)$.

\begin{figure}[t]
\begin{center}
\includegraphics[width=\columnwidth]{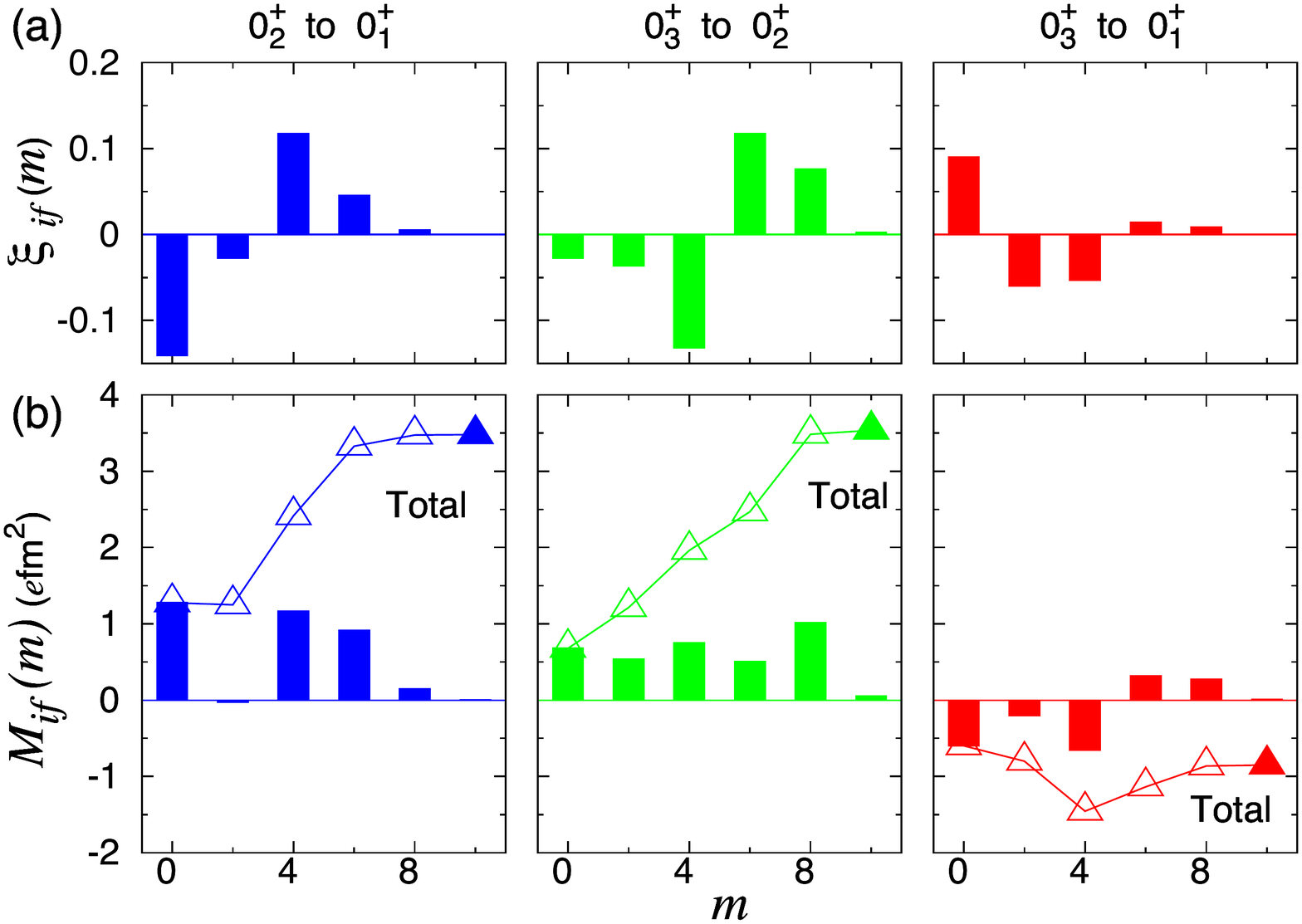}
\vspace*{-0.7cm}
\end{center}
\captionsetup[figure]{font={small,stretch=1.0}}
\caption{\label{fig:dm0_hw} (a) Overlap $\xi_{if}(m)$ and (b) $E0$ matrix element $M_{if}(m)$ for
 each $m$ value of the $mp$-$mh$ excitation.
 The summed $E0$ matrix elements $\sum_{m'\le m} M_{if}(m')$ and the total values are shown by the
 open and filled triangles, respectively.}
\end{figure}

In Figure~\ref{fig:dm0_hw}, $\xi_{if}(m)$ and $M_{if}(m)$ are plotted
for each contributing value of $m$. The orthogonality condition 
also allows the weighting factors to be adjusted without affecting
the total $E0$ matrix element. In this case, the large values of $w(m)$
for $^{40}$Ca ($m + 120$) may be replaced with a smaller value, $w(m') =
m - \overline{m} $, where $ \overline{m} = \sum_m m \times \xi(m)^2 /
\sum_m \xi(m)^2 $ for each transition. 
The total $E0$ strength is independent of the choice of $\overline{m}$, 
but such a shift reduces the amplitude of each $M_{if}(m)$ value, 
thus facilitating the understanding
of how the individual mixing amplitudes contribute to the total $E0$
matrix element. The $M_{if}(m)$ values shown in Fig.~\ref{fig:dm0_hw}(b)
were determined with the scaled weighting factors, $w(m')$.

Figure~\ref{fig:dm0_hw}(a) shows that both the $0^+_{2} \to 0^+_{1}$ and
$0^+_{3} \to 0^+_{2}$ transitions have negative $\xi_{if}(m)$ values for
small $m$ that change sign once and become positive at larger $m$. In
such cases, $w(m')$ may be chosen so that the $M_{if}(m)$ values add
constructively, as presented in Fig.~\ref{fig:dm0_hw}(b), and the
resulting $E0$ matrix element is moderate. On the other hand, the
$0^+_{3} \to 0^+_{1}$ transition is rather different: the $\xi_{if}(m)$
values change sign twice at $m=0 \to 2$ and $m=4\to 6$. In this case,
destructive interference of $M_{if}(m)$ is inevitable for any possible
value of $w(m')$.

Finally, the two-state description is revisited in terms of the present
microscopic analysis. A system obeying this model should have two
non-vanishing $\xi_{if}(m)$ values at $m=m_1, m_2$ constrained to
$\xi_{if}(m_1) + \xi_{if}(m_2)=0$. This system has only one degree of
freedom, corresponding to $\alpha$ in Eq.~(\ref{eq:two_state}); no phase
factor is relevant to the observables. In contrast, when three states
are involved there are two degrees of freedom, and interference occurs
between them. The very small $\rho^2(E0;0^+_3\to 0^+_1)$ value is
a manifestation of this new aspect of shape coexistence with
three-state mixing.

Note that the degree of the mixing between SD and ND can be
probed with $B(E2;0^+_3 \to 2^+_1)$.
The calculated value, $1.3$~W.u., is significantly smaller than the measured
value, $17^{+4}_{-3}$~W.u.
Considering the moderate $\rho(E0;0^+_3 \to 0^+_2)$ value by LSSM (see Fig.~3),
we do not expect larger mixing between those bands.
This underestimate may be due to incorrect $K$ mixing between
the $2^+_1$ and $2^+_2$ states.

In summary, the $E0$ transition strength of $\rho^2(E0; 0^+_3 \to 0^+_1)
= 2.3(5)\times10^{-3}$ was measured for the first time in doubly magic
$^{40}$Ca; this is the lowest $\rho^2(E0)$ value measured among nuclei
of $A<50$. An upper limit of $\rho^2(E0; 0_3^+ \to 0_2^+) <
4.5\times10^{-2}$ was also obtained. Large-scale shell-model
calculations were performed for $^{40}$Ca and several neighboring
nuclei. The calculated $E0$ matrix elements were analyzed in terms of
multiparticle-multihole configuration mixing. Moderately large
$\rho^2(E0)$ values for the $0^+_2 \to 0^+_1$ and $0^+_3 \to 0^+_2$
transitions are consistent with a two-state model, in which only a
squared mixing amplitude matters. Conversely, the extremely small
$\rho^2(E0;0^+_3 \to 0^+_1)$ value is caused by destructive interference
among $mp$-$mh$ components in a three-state mixing scenario. The new
data and shell-model calculations provide a novel perspective on
multiple shape coexistence, with implications for experimental and
theoretical activities extending to nuclei far from stability.

 This work is supported by the International Joint Research Promotion
 Program of Osaka University, JSPS KAKENHI Grant Number JP 17H02893, 
 18H03703, JSPS A3 Foresight program Grand Number JPJSA3F20190002, and
 the Australian Research Council grant numbers DP140102896 and
 DP170101673. N.~S. and Y.~U. acknowledge KAKENHI grants (20K03981,
 17K05433), ``Priority Issue on post-K computer'' (hp190160, hp180179,
 hp170230) and ``Program for Promoting Researches on the Supercomputer
 Fugaku'' (hp200130, hp210165), MEXT, Japan. E.I., N.A., N.S., and
 Y.U. acknowledge support from the RCNP Collaboration Research Network
 (RCNP COREnet); A.A., B.J.C., J.T.H.D., T.J.G., and B.P.M. acknowledge
 support of the Australian Government Research Training Program. 
 Support for the ANU Heavy Ion Accelerator Facility
 operations through the Australian National Collaborative Research
 Infrastructure Strategy (NCRIS) program is also acknowledged.
 The authors thank J.~Heighway for preparing targets.
E.I. acknowledges fruitful discussions with J.L.~Wood on issues of
shape coexistence.


\begin{thebibliography}{50}

\bibitem{Hor06} M.~Horoi and K.~A.~Jackson,
     {Chem. Phys. Lett. \textbf{427}, 147 (2016)}
      and references therein.
\bibitem{Lag04} ``Molecular Structure'' Chemistry Foundations and
	Applications. Volume 3. Farmington, MI: J.J.~Lagowski, 2004. 

\bibitem{hey11} K.~Heyde and J.~L.~Wood,
      {Rev. Mod. Phys. \textbf{83}, 1467 (2011).}
\bibitem{Tan19} R.~Taniuchi, \emph{et al.},
      {Nature \textbf{569}, 53 (2019).}
\bibitem{Gui84} D.~Guilemaud-Mueller, \emph{et al.},
      {Nucl. Phys. A \textbf{426}, 37 (1984).}
\bibitem{Mot95} T.~Motobayashi, \emph{et al.},
      {Phys. Lett. B \textbf{346}, 9 (1995).}
\bibitem{Wim10} K.~Wimmer, T.~Kr\"oll, R.~Kr\"ucken, V.~Bildstein,
	R.~Gernh\"auser, B.~Bastin, \emph{et al.},
      {Phys. Rev. Lett. \textbf{105}, 252501 (2010).}
\bibitem{For11} H.~T.~Fortune,
      {Phys. Rev. C \textbf{84}, 024327 (2011).}
\bibitem{For12} H.~T.~Fortune,
      {Phys. Rev. C \textbf{85}, 014315 (2012).}
\bibitem{Mac16} A.~O.~Macchiavelli, H.L.~Crawford, C.M.~Campbell,
	R.M.~Clark, M.~Cromaz, P.~Fallon, \emph{et al.},
       {Phys. Rev. C \textbf{94}, 051303(R) (2016).}
\bibitem{Mac17} A.~O.~Macchiavelli and H.~L.~Crawford,
       {Phys. Scr. \textbf{92}, 064001 (2017).}
\bibitem{Had18} K.~Hady\ifmmode \acute{n}\else \'{n}\fi{}ska-Kl\ifmmode \mbox{\c{e}}\else \c{e}\fi{}k,
 \emph{et al.},
       {Phys. Rev. C \textbf{97}, 024326 (2018).}
\bibitem{ide01} E.~Ideguchi, D.G.~Sarantites, W.~Reviol, A.V.~Afanasjev,
	M.~Devlin, C.~Baktash, \emph{et al.},
      {Phys. Rev. Lett. \textbf{87}, 222501 (2001).}

\bibitem{chi03} C.J.~Chiara, E.~Ideguchi, M.~Devlin, D.R.~LaFosse, F.~Lerma, 
	W.~Reviol, S.K.~Ryu, D.G.~Sarantites, \emph{et al.},
	{Phys. Rev. C \textbf{67}, 041303(R) (2003).}
\bibitem{ger69} W.~J.~Gerace and A.~M.~Green,
      {Nucl. Phys. A \textbf{123}, 241 (1969).}

\bibitem{For79} H.T.~Fortune, M.N.I.~Al-Jadir, R.R.~Betts, J.N.~Bishop,
	R.~Middleton, 
	{Phys. Rev. C \textbf{19}, 756 (1979).}

\bibitem{Mid72} R.~Middleton, J.~Garrett, H.T.~Fortune,
        {Phys. Lett. B \textbf{39}, 339 (1972).}

\bibitem{Boh77} W.~Bohne, K.D.~Buchs, H.~Fuchs, \emph{t al.},
	{Nucl. Phys. A \textbf{284}, 14 (1977).}

%
\bibitem{Ger67} W.J.~Gerace and A.M.~Green,
	{Nucl. Phys. A \textbf{93}, 110 (1967).}
	
\bibitem{cau07} E.~Caurier, J.~Men\'{e}ndez, F.~Nowacki, A.~Poves, 
       {Phys. Rev. C \textbf{75}, 054317 (2007).}
\bibitem{sin02} B.~Singh, R.~Zywina, R.~B.~Firestone,
      {Nucl. Data Sheets \textbf{97}, 241 (2002).}
\bibitem{sve00} C.~E.~Svensson, A.O.~Macchiavelli, A.~Juodagalvis,
	A.~Poves, I.~Ragnarsson, S.~{\AA}berg, \emph{et al.},
  {Phys. Rev. Lett. \textbf{85}, 2693 (2000).}
\bibitem{ide10} E.~Ideguchi, \emph{et al.},
       {Phys. Lett. B \textbf{686}, 18 (2010).}
%
\bibitem{Ald72} A.V.~Aldushchenkov and N.A.~Voinova, 
{Nucl. Data Tables \textbf{11}, 299 (1972).}

\bibitem{woo99} J.~L.~Wood \emph{et al.},
      {Nucl. Phys. A \textbf{651}, 323 (1999).}

\bibitem{gar16} R.~F.~Garcia~Ruiz, \emph{et al.},
       {Nature Phys. \textbf{12}, 594 (2016).}

\bibitem{ulr77} M.~Ulrickson, \emph{et al.},
       {Phys. Rev. C \textbf{15,} 186 (1977).}
\bibitem{kib90} T.~Kib\'{e}di, G.~D.~Dracoulis and A.~P.~Byrne,
       {Nucl. Instrum. Meth. A \textbf{294}, 523 (1990).}
\bibitem{Eri20} T.~K.~Eriksen, T.~Kib\'{e}di, M.W.~Reed, A.E.~Stuchbery,
	K.J.~Cook, A.~Akber, \emph{et al.},
        {Phys. Rev. C \textbf{102}, 024320 (2020).}
\bibitem{Dow20_PLB} J. T. H. Dowie, \emph{et al.},
        {Phys. Lett. B \textbf{811}, 135855 (2020).}
       {Nucl. Phys. A \textbf{131}, 113 (1969).}

\bibitem{nds140} J.~Chen,
       {Nucl. Data Sheets \textbf{140}, 1 (2017).}
\bibitem{til93} D.~R.~Tilley, H.~R.~Weller, and C.~M.~Cheves,
       {Nucl. Phys. A \textbf{564}, 1 (1993).}
\bibitem{do20} J.~T.~H.~Dowie, \emph{et al.}, 
       {At. Data Nucl. Data \textbf{131}, 101283 (2020).}
\bibitem{kib08} T.~Kib\'{e}di, \emph{et al.},
       {Nucl. Instrum. Meth. A \textbf{589}, 202 (2008).}
\bibitem{pol69} A.~R.~Poletti, \emph{et al.},
       {Phys. Rev. \textbf{181}, 1606 (1969).}
\bibitem{kib22} T.~Kib\'{e}di, A.B.~Garnsworthy, and J.L.~Wood,
      {Prog. Part. and Nucl. Phys. \textbf{123}, 103930 (2022)}



\bibitem{uts99} Y.~Utsuno, T.~Otsuka, T.~Mizusaki and M.~Honma,
       {Phys. Rev. C \textbf{60}, 054315 (1999).}
\bibitem{uts04} Y.~Utsuno, T.~Otsuka, T.~Glasmacher, T.~Mizusaki,
	M.~Honma, 
       {Phys. Rev. C \textbf{70}, 044307 (2004).}
\bibitem{go21} S.~Go, E.~Ideguchi, R.~Yokoyama, N.~Aoi, F.~Azaiez,
	K.~Furutaka, \emph{et al.},
       {Phys. Rev. C \textbf{103}, 034327 (2021).}
\bibitem{oln71} J.~W.~Olness, W.~R.~Harris, A.~Gallmann, \emph{et al.},
       {Phys. Rev. C \textbf{3}, 2323 (1971).}
\bibitem{wood92} J.~L.~Wood, \emph{et al.}, 
       {Phys. Rep. \textbf{215}, 101 (1992).}
\bibitem{had16} K.~Hady\ifmmode \acute{n}\else \'{n}\fi{}ska-Kl\ifmmode \mbox{\c{e}}\else \c{e}\fi{}k,
 \emph{et al.},
        {Phys. Rev. Lett. \textbf{117}, 062501 (2016).}
%

\end{thebibliography}
\end{document}